\documentclass[12pt]{article}
\usepackage{epsfig}
\usepackage{amsfonts}
\usepackage{latexsym}
\usepackage{amsmath}
\usepackage{mathrsfs}
\usepackage{hyperref}
\numberwithin{equation}{section}

\DeclareFontFamily{OT1}{rsfs}{}
\DeclareFontShape{OT1}{rsfs}{m}{n}{
<-7> rsfs5 <7-10> rsfs7 <10-> rsfs10}{}
\DeclareMathAlphabet{\mycal}{OT1}{rsfs}{m}{n}

\def\half{{1\over 2}}
\newcommand{\p}{\partial}

\newcommand{\bea}{\begin{eqnarray}}
\newcommand{\eea}{\end{eqnarray}}
\newcommand{\be}{\begin{equation}}
\newcommand{\ee}{\end{equation}}

  \makeatletter
  \let\over=\@@over \let\overwithdelims=\@@overwithdelims
  \let\atop=\@@atop \let\atopwithdelims=\@@atopwithdelims
  \let\above=\@@above \let\abovewithdelims=\@@abovewithdelims
  \makeatother

\begin{document}

\begin{titlepage}
\unitlength = 1mm
\begin{center}

{ \LARGE {\textsc{Gravity Waves from Kerr/CFT}}}

\vspace{0.8cm}
Achilleas P. Porfyriadis and Andrew Strominger

\vspace{1cm}

{\it  Center for the Fundamental Laws of Nature, Harvard University,\\
Cambridge, MA 02138, USA}

\begin{abstract}
Dynamics at large redshift near the horizon of an extreme Kerr black hole are governed by an infinite-dimensional conformal symmetry. This symmetry may be exploited to analytically, rather than numerically, compute a variety of potentially observable processes. In this paper we compute and study the conformal transformation properties of the gravitational radiation emitted by an orbiting mass in the large-redshift near-horizon region.
\end{abstract}

\vspace{1.0cm}

\end{center}
\end{titlepage}

\pagestyle{plain}
\setcounter{page}{1}
\newcounter{bean}
\baselineskip18pt


\setcounter{tocdepth}{2}
\tableofcontents
\section{Introduction}

Astronomical observation suggests the existence of near-extreme Kerr black holes whose horizons spin at nearly the speed of light.  Examples  include the nearby stellar mass black holes GRS 1915+105 \cite{McClintock:2006xd} and Cygnus X-1 \cite{Gou:2013dna}, and the supermassive black hole at the center of the Seyfert-1.2 galaxy MCG-6-30-15 \cite{Brenneman:2006hw}. General relativity implies \cite{hep-th/9905099, 0809.4266} that the dynamics of the high-redshift near-horizon region of extreme Kerr, which includes the innermost-stable-circular-orbit (ISCO), is governed by an infinite-dimensional emergent conformal symmetry. Symmetries of physical systems in general may both usefully characterize and have striking consequences for observational data.

Precision black hole spectroscopy has advanced to the stage where we are beginning to observe - with a variety of ingenious methods - the regions of spacetime governed by this conformal symmetry. A primary purpose of this paper is to embark on an exploration of potential phenomenological consequences of the symmetry. In particular, we  will study the gravitational radiation produced by a massive test object orbiting very near the horizon of an extreme Kerr black hole.
The near-horizon approximation is valid, and the conformal symmetry relevant, for example for a compact stellar mass object orbiting within $10^{-2}$ au of the supermassive black hole in MCG-6-30-15.  The resulting radiation is potentially observable at eLISA \cite{Finn:2000sy,Gair:2004iv,elisa}.
Such extreme-mass-ratio-inspirals (EMRIs) have been extensively studied with a variety of methods; see  e.g. \cite{Hughes:1999bq, Hughes:2001jr, Mino:2003yg, Hinderer:2008dm, Hadar:2009ip, Pan:2010hz, Yunes:2010zj,Zhang:2011vha} for recent progress and \cite{Sasaki:2003xr, AmaroSeoane:2007aw, Blanchet:2013haa} for recent reviews. Previous computations of the gravitational waveform typically employ either the post-Newtonian expansion or numerical methods, both of which face important challenges in the interesting case of extreme Kerr. Here, we exploit the conformal symmetry and develop a complementary analytic approach.  One hopes that the greatly enhanced symmetry will enable an analytic treatment of a variety of astrophysical phenomena in the near-horizon region of extreme Kerr,  such as  self-force \cite{Barack:2009ux, Poisson:2011nh} or transient resonances \cite{Flanagan:2010cd}.

In this paper we consider only the case of an extreme Kerr black hole. The near-extreme case is treated, to leading order in the deviation from extremality, in the companion paper \cite{Hadar:2014dpa}. Massive objects orbiting a near-extreme Kerr black hole quickly plunge into the horizon after passing the ISCO. Computing the radiation production during the plunge phase appears much more difficult, but the plunge is still soluble because it is related by a conformal transformation to the circular orbits of this paper. In the plunge context the utility of the conformal symmetry comes into full play.

The emergence of conformal symmetry at large redshifts near an extreme Kerr horizon is mathematically similar to the more familiar emergence of conformal symmetry at low energies in condensed matter systems near a critical point. This leads to many striking observed phenomena: for example critical opalescence at the liquid-gas transition of carbon dioxide. In the astrophysical setting, it also resembles the emergence of conformal symmetry at large cosmological redshifts posited in the theory of inflation. This implies the observed scale invariance of the CMB spectrum. Although we do not propose a candidate smoking gun for the conformal symmetry of extreme Kerr herein, we hope that future investigations will produce one.

In a second motivation, the present work also bears on ongoing investigations in quantum gravity. The extreme Kerr horizon bears similarities to that of certain supersymmetric string theoretic black holes, for which it is known \cite{Strominger:1996sh, Maldacena:1996ix, hep-th/9702015,Maldacena:1997re} that, at low energies and large redshifts, the near horizon dynamics has a completely equivalent ``dual'' description as a two-dimensional conformal field theory (CFT). In this case the near-horizon geometry has an anti-deSitter (AdS) factor, and this equivalence is an early example of the AdS/CFT correspondence. The close resemblance of the stringy and Kerr horizons motivated the ``Kerr/CFT conjecture'' which posits that quantum gravity near the extreme Kerr horizon is also dual to a two-dimensional boundary CFT \cite{0809.4266} (see \cite{Bredberg:2011hp,Compere:2012jk} for reviews). A secondary purpose of this paper is to both test and illuminate the Kerr/CFT conjecture through explicit computation. We shall see that the presence of an orbiting star induces a computable deformation of the CFT, and the gravitational radiation then follows from the application of Fermi's golden rule to the deformed CFT. The perfect and detailed agreement\footnote{Perhaps even better than expected, see section 3.3 for discussion.} between the bulk gravity and boundary CFT computations of the radiation rate found herein provide new evidence for the conjecture.
Although mathematically far less understood than the AdS/CFT conjecture, the Kerr/CFT conjecture pertains to observable regions of our universe. The classical conformal symmetry on which it is based may,  with future advances in both theory and observation,  ultimately be directly observed.

This paper is organized as follows. Section 2 contains a telegraphic review of the Kerr metric, the extreme and near-horizon  limits,  the decoupled Near-Horizon Extreme Kerr  - or NHEK - geometry and Kerr/CFT. In section 3 we introduce, as a warmup, a scalar field coupled to the orbiting star. Scalar radiation is computed for both the decoupled NHEK and the full asymptotically flat case. The NHEK computation is shown to agree with the dual computation for the boundary CFT. In section 4 we adopt the Newman-Penrose formalism and compute gravitational radiation for both the asymptotically flat and the NHEK case. Agreement is again found with the application of Fermi's golden rule to a deformation of the boundary CFT.

\section{Review of Kerr, NHEK and Kerr/CFT}
This section contains a lightning review of Kerr, its NHEK limit, and the Kerr/CFT correspondence.

The metric of the Kerr black hole in Boyer-Lindquist coordinates is ($G=\hbar=c=1$):
\be\label{Kerr}
ds^2=-{\Delta \over\hat \rho^2}\left(d\hat t-a \sin^2\theta d\hat\phi\right)^2+{\sin^2 \theta \over \hat \rho^2}
\left((\hat r^2+a^2)d\hat \phi-a d\hat t\right)^2+{\hat\rho^2 \over\Delta}d\hat r^2+\hat \rho^2 d\theta^2\,,
\ee
\be
\Delta=\hat r^2-2M\hat{r}+a^2\,,\quad \hat \rho^2=\hat r^2 +a^2\cos^2 \theta \,. \notag
\ee
It is labeled by two parameters: the mass $M$ and angular momentum $J = a M$. We consider the extreme limit in which
\be
J=M^2\,,\quad a=M\,.
\ee
The horizon is then at $\hat r=M$ and the proper spatial distance to it is infinite, which, as shown by Bardeen and Horowitz \cite{hep-th/9905099}, allows us to zoom in on the near horizon region and treat it as its own spacetime. We may take the
near horizon limit by defining
\be\label{coord transn}
t = {\hat{t}\over 2M}\,, \quad r = {\hat{r} - M\over  M}\,,\quad \phi = \hat{\phi}-{\hat{t}\over 2M}\,,
\ee
and considering the small $r$ limit. The resulting NHEK geometry is:
\be\label{NHEK}
ds^2 = 2M^2 \Gamma(\theta)\left( -r^2 dt^2 + {dr^2 \over r^2} + d\theta^2 + \Lambda(\theta)^2(d\phi + r dt)^2\right)\,,
\ee
where
\be
\Gamma(\theta) = {1+\cos^2\theta\over 2} \ , \quad \Lambda(\theta) = {2\sin\theta\over 1 + \cos^2\theta}\,,
\ee
and $\phi \sim \phi + 2\pi\,, 0 \leq \theta \leq \pi$.  The entrance to the throat, where the near horizon region glues onto the full asymptotically flat Kerr geometry, corresponds to the boundary of NHEK at $r=\infty$. NHEK has an enhanced isometry group,
\be
U(1)_L \times SL(2,\mathbb{R})_R\,,
\ee
generated by the Killing vectors:
\bea
Q_0&=&-\p_\phi\,,\cr
H_{-1}&=&\p_t\,,\cr
H_0&=&t\p_t-r\p_r\,,\cr
H_1&=&{1 \over 2r^2}\p_t+{t^2\over 2}\p_t-tr\p_r-{1 \over r}\p_\phi\,.
\eea
Excitations of NHEK carry the associated charges. These isometries govern the dynamics of NHEK in the sense that all processes must lie in representations of $U(1)_L \times SL(2,\mathbb{R})_R$. However, it is often the case in general relativity \cite{Brown:1986nw,Porfyriadis:2010vg} that the diffeomorphisms which act nontrivially and govern the physical dynamics - known as the asymptotic symmetry group or ASG - are comprised of  more than the isometries. In \cite{0809.4266} it was shown, for a certain choice of boundary conditions enforcing $M^2=J$, that the ASG contains a full Virasoro symmetry whose zero mode is the angular momentum $Q_0 =-\p_\phi$. Non-extremal excitations with $M^2 - J > 0$ correspond to non-zero charges under $SL(2,\mathbb{R})_R$. Boundary conditions allowing for such excitations were introduced in \cite{Matsuo:2009sj,Matsuo:2009pg,Rasmussen:2009ix} and found to lead to a second Virasoro in the ASG whose zero mode is the energy.

The appearance of Virasoro symmetries in the classically computed ASG, together with the analogy to AdS$_3$/CFT$_2$ \cite{Brown:1986nw,Strominger:1996sh, Maldacena:1996ix, hep-th/9702015,Maldacena:1997re} motivates the conjecture that bulk quantum gravity on the NHEK region of Kerr is dual to a two-dimensional boundary CFT.\footnote{The CFT duals to NHEK are expected to be more exotic than their AdS$_3$ counterparts, possibly involving non-local or warped behavior and reflecting the superradiant instabilities \cite{Hofman:2011zj, ElShowk:2011cm, Song:2011sr, Bena:2012wc, Detournay:2012pc, Guica:2013jza}.} Reviews can be found in \cite{Bredberg:2011hp, Compere:2012jk}. From the first law of black hole thermodynamics one concludes that the CFT must be at the left-moving temperature (conjugate to $Q_0$) $T_L ={1/2\pi}$. Conformal symmetry can be used to derive a dictionary relating bulk gravity quantities to their CFT counterparts. The conjecture predicts that all processes in NHEK can be mapped to and computed as a process in the dual finite-temperature two-dimensional CFT. We shall verify herein that this is indeed the case for gravity wave production by an orbiting star.
We wish to stress that the Kerr/CFT conjecture goes beyond the existence of classical conformal symmetries, whose consequences are the primary focus of this paper.

\section{Scalar radiation from a star orbiting near the horizon}
In this section we consider a test particle - or ``star'' - in an eternal (relative to asymptotic time) circular orbit in the NHEK geometry. We couple it to a massless scalar field and compute and compare the resulting scalar radiation at the horizon from the bulk gravity and boundary CFT perspectives. We then attach the asymptotically flat region back and compute the radiation flux at future null infinity as well.
\subsection{Gravity analysis in NHEK}

Consider a  star orbiting at radius $r_0$ in the NHEK geometry (\ref{NHEK}). We may parameterize the corresponding geodesic $x_*^\mu(t)$ with the NHEK time $t$:
\begin{eqnarray}\label{plunge in NHEK}
x_*^t(t)&=&t\,,\cr
x_*^\phi(t)&=&\phi_0-\frac{3}{4}r_0\,t\,, \cr
x_*^r(t)&=&r_0\,,\cr
x_*^\theta(t)&=&{\pi \over 2}\,.
\end{eqnarray}
This has NHEK energy and angular momentum (per unit rest mass)
\be
E=-g_{t\mu}\p_\tau x_*^\mu=0\,,\quad L=g_{\phi \mu}\p_\tau x_*^\mu={2M\over \sqrt{3}}\,,
\ee
where $\tau$ is the proper time.

We couple the star  to a massless scalar field $\Psi$ with the interaction
\be\label{Sint}
S_I=4\pi\lambda \int d\tau \, \Psi(x_*(\tau))\,,
\ee
where $\lambda$ is a coupling constant.
The scalar wave equation in the presence of the star is then:
\be\label{wave eqn}
\square\, \Psi=4\pi {\cal T}\,,
\ee
where (setting  $\phi_0=0$)
\be\label{T for NHEK plunge}
{\cal T}=-\frac{\sqrt{3}\lambda r_0}{4M^3}\delta(r-r_0)\delta(\theta-{\pi\over 2})\delta(\phi+{3\over 4}r_0 t)\,.
\ee
The scalar source preserves one Killing symmetry:
\be\label{KIlling vector}
\chi=\p_t-{3\over 4}r_0\,\p_\phi\,.
\ee
A $\chi$-invariant solution to the wave equation may be constructed using the mode expansion
\begin{eqnarray}
\Psi&=&\sum_{\ell,m}e^{im\left(\phi+3r_0t/4\right)}S_\ell(\theta) R_{\ell m}(r)\,,\label{psi simpler  expansion}\\
\,{\cal T}&=&{1 \over 8 \pi M^2\Gamma(\theta)}\sum_{\ell,m}e^{im\left(\phi+3r_0t/4\right)}S_\ell(\theta) T_{\ell m}(r)\,,\label{source simpler expansion}
\end{eqnarray}
where $S_\ell$ are the spheroidal harmonics obeying
\be {1\over\sin\theta}\p_\theta(\sin\theta\,\p_\theta S_\ell)+\left(K_\ell-\frac{m^2}{\sin^2\theta}-\frac{m^2}{4}\sin^2\theta\right)S_\ell= 0\,,\label{angular eqn}
\ee
with $K_\ell$ a separation constant, $\ell\ge 0\,, -\ell\le m\le \ell$.  $S_\ell$ and $K_\ell$ depend on both $m$ and $\ell$ but we write only their primary label $\ell$ to avoid index clutter. The spheroidal harmonics are normalized as:
\be\label{sph harmonics normn}
\int_0^\pi\sin\theta d\theta\,S_\ell(\theta) S_{\ell'}(\theta)=\delta_{\ell \ell'}\,.
\ee
The expansion coefficients for ${\cal T}$ are
\begin{eqnarray}
T_{\ell m}(r)
&=&4M^2\int d\left(\phi+3r_0t/4\right)\,\sin\theta d\theta\, e^{-im(\phi+3r_0\, t/4)}S_\ell(\theta)\Gamma(\theta)\, {\cal T} \nonumber\\
&=&-\frac{\sqrt{3}\lambda r_0}{2M}S_\ell(\pi/2)\delta(r-r_0)\,.  \label{T_lm simple}
\end{eqnarray}
The separated radial equation becomes
\be
\p_r(r^{2}\p_r R_{\ell m})+\left(2m^2-K_\ell+\frac{2\omega m}{r}+\frac{\omega^2}{r^2}\right)R_{\ell m}=T_{\ell m}\,,\label{radial eqn}
\ee
where
\be\label{Omega defn}
\omega=-{3\over 4}mr_0\,.
\ee

The radial equation is a Sturm-Liouville problem that may be solved, for given boundary conditions, via standard Green function methods as follows. Two linearly independent solutions to the homogeneous radial equation ($T_{\ell m}=0$) are given by the Whittaker (i.e. confluent hypergeometric) functions \cite{0906.2376}:
\be\label{Whittaker solns}
M_{im,h-\half}\left(-{2i\omega}/{r}\right)\,,\quad W_{im,h-\half}\left(-{2i\omega}/{r}\right)\,,
\ee
where
\be\label{h defn}
h\equiv\half+\sqrt{1/4+K_\ell-2m^2}\,.
\ee
In this paper, we will only consider the case
\be
\mathrm{Re}[h]<1\,.
\ee
Then the Whittaker functions have the following asymptotic behaviors:
\begin{eqnarray}
M_{im,h-\half}\left(-{2i\omega}/{r}\right) &\to& A\, r^{-im} e^{i\omega/r} +B\, r^{im} e^{-i\omega/r} \quad \textrm{for $r\to 0$}\,,\nonumber\\
 &\to& (-2i\omega)^{h}\, r^{-h} \qquad\qquad\qquad~~\, \textrm{for $r\to \infty$}\,,\\
W_{im,h-\half}\left(-{2i\omega}/{r}\right) &\to& (-2i\omega)^{im}\, r^{-im} e^{i\omega/r} \qquad\qquad \textrm{for $r\to 0$}\,,\nonumber\\
 &\to& C\, r^{h-1}+D\, r^{-h} \qquad\qquad\quad~\, \textrm{for $r\to \infty$}\,,\label{say}
\end{eqnarray}
where
\begin{eqnarray}
&&A=(2i\omega)^{im}\frac{(-i\omega)^{h}}{(i\omega)^{h}}\frac{\Gamma(2h)}{\Gamma(h+im)}\,,
\quad B=(-2i\omega)^{-im}\frac{\Gamma(2h)}{\Gamma(h-im)}\,, \label{Abar, Bbar expressions} \\
&&C= (-2i\omega)^{1-h}\frac{\Gamma(2h-1)}{\Gamma(h-im)}\,,
\qquad D=(-2i\omega)^{h}\frac{\Gamma(1-2h)}{\Gamma(1-h-im)}\,. \label{Cbar, Dbar expressions}
\end{eqnarray}
Note that $W_{im,h-\half}\left(-{2i\omega}/{r}\right)$ is purely ingoing at the horizon, but has both falloffs at infinity. $M_{im,h-\half}\left(-{2i\omega}/{r}\right)$ on the other hand has both incoming and outgoing modes at the horizon but only one falloff, $r^{-h}$, at infinity,
which we will refer to as the Neumann boundary condition.

To go further we must specify boundary conditions on the solution. We consider the solution of \eqref{radial eqn} that is purely ingoing at the horizon and obeys Neumann boundary conditions at $r=\infty$. This is given by:
\be\label{radial soln}
R_{\ell m}(r)={1\over W}\left[X\,\Theta(r_0-r) W_{im,h-\half}\left(-{2i\omega}/{r}\right) +Z\,\Theta(r-r_0) M_{im,h-\half}\left(-{2i\omega}/{r}\right)\right]
\ee
where $W$ is the $r$-independent Wronskian,
\be
W=r^2\left[W_{im,h-\half}\left(-{2i\omega}/{r}\right) \p_r M_{im,h-\half}\left(-{2i\omega}/{r}\right) - \left(W\leftrightarrow M\right) \right] =2i\omega\frac{\Gamma(2h)}{\Gamma(h-im)}\,,
\ee
and
\be
X=-\frac{\sqrt{3}\lambda r_0}{2M} S_\ell(\pi/2)  M_{im,h-\half}\left({3im}/{2}\right)\,,\quad
Z=-\frac{\sqrt{3}\lambda r_0}{2M}S_\ell(\pi/2) W_{im,h-\half}\left({3im}/{2}\right)\,.
\ee
Putting everything together we have:
\begin{eqnarray}
\Psi(r\to 0)&=&\sum_{\ell,m} e^{im(\phi+3r_0t/4)}\,S_\ell(\theta)\,\frac{ X}{W} (-2i\omega)^{im}\, r^{-im}e^{-3imr_0/4r}\,,\\
\Psi(r\to \infty)&=&\sum_{\ell,m}e^{im(\phi+3r_0t/4)}\,S_\ell(\theta)\,\frac{ Z}{W} (-2i\omega)^{h} \,r^{-h}\,.
\end{eqnarray}
The Klein-Gordon particle number flux,
\be
{\cal F}=-\int\sqrt{-g}J^r d\theta d\phi\,,\quad J^\mu={i\over 8\pi}(\Psi^*\nabla^\mu\Psi-\Psi\nabla^\mu\Psi^*)\,,\quad \nabla_\mu J^\mu=\mathrm{Im}[\Psi {\cal T}]\,,
\ee
vanishes at infinity for real $h$, while at the horizon we have:
\be\label{particle number flux}
{\cal F}_{\ell m}={\lambda^2\over 4} r_0 m^{-1}e^{-\pi m}S_\ell^2(\pi/2)\frac{\left|\Gamma(h+im)\right|^2}{|\Gamma(2h)|^2} \left|M_{im,h-\half}\left({3im}/{2}\right)\right|^2\,,
\ee
for $m>0$.

To summarize, the orbiting star emits scalar radiation. This radiation is reflected off of the NHEK boundary with Neumann boundary conditions, and falls into the black hole. The particle number flux across the future horizon is given by (\ref{particle number flux}).

\subsection{CFT analysis}

Let us now consider the dual description of the scalar emission by an orbiting star  as a process in the two-dimensional boundary CFT. Analogous types of dualities have been analyzed in the context of AdS/CFT \cite{hep-th/9809188,hep-th/9808017, hep-th/9812007}.
The effect of the star at fixed radius $r_0$ in NHEK may be compared to that of, e.g.,  adding a static $D_3$ brane in AdS$_5$ with flux $N$ at fixed radius $r_0$ in Poincare coordinates while maintaining the asymptotic boundary conditions. In the AdS$_5$ case the resulting geometry is unaffected for $r>r_0$ but corresponds to flux $N-1$ for $r<r_0$. The radius $r_0$ is identified with a scale in the dual $N=4$ $U(N)$ gauge theory.  Processes above this scale, or outside the D-brane, are governed by $U(N)$ gauge theory, while those at low scales inside the D-brane are governed by $U(N-1)$ gauge theory.

In our example the orbiting star should lead to a deformed CFT at scales corresponding to $r<r_0$. To find that CFT, we extend the solution in the region $r<r_0$ all the way out to the boundary at $r=\infty$ ignoring the discontinuity produced by the star. This extended solution will contain both Neumann and Dirichlet modes at infinity. The CFT deformation is then read off of the coefficients of the Dirichlet modes.\footnote{One may alternately, for $\mathrm{Re}[h]<1$, define the deformation in terms of the Neumann modes: this would be dual to the gravity problem with Dirichlet boundary conditions \cite{Klebanov:1999tb}.} 

To be explicit, we write the action of the deformed CFT as
\be\label{pt}
S = S_{CFT}+\sum_{\ell}\int dt^+dt^-J_{\ell}(t^+,t^-)\mathcal{O}_{\ell}(t^+,t^-)\,.
\ee
Here $S_{CFT}$ is the original CFT action and $J_{\ell}$ are to-be-determined c-number source functions, and $\mathcal{O}_{\ell}(t^+,t^-)\equiv\sum_m \mathcal{O}_{\ell m}(t^-)e^{-imt^+}$. The operators $\mathcal{O}_{\ell m}$ are the boundary duals to the mode of the bulk scalar with separation constant $K_\ell$ and angular momentum $m$. According to the Kerr/CFT dictionary (see e.g. \cite{0907.3477}) these have left and right conformal weight:
\be
h_L=h_R=h\,.
\ee
In this subsection, we consider only real $h$.
The bulk isometries $\p_\phi$ and $\p_t$ are identified, up to a scale, with left and right translations in the 2D CFT, which implies
\be t^+=\phi\,,\quad t^-= t\,.\ee
It follows from the  symmetry \eqref{KIlling vector} that $J_{\ell}(t^+, t^-)$ can only depend on the combination $t^++3r_0t^-/4$. We accordingly Fourier expand:
\be
J_{\ell}(t^+,t^-) = \sum_{m}J_{\ell m}e^{im(t^++3r_0t^-/4)}\,.
\ee
$\phi$-periodicity implies that $m$ is an integer.
As discussed above, the coefficients $J_{\ell m}$ are then read off the extension $R^{ext}_{\ell m}$ of the small $r$ behavior of the radial solution $R_{\ell m}$ in (\ref{radial soln}),
\be
R^{ext}_{\ell m}(r)={ X\over W}  W_{im,h-\half}\left(-{2i\omega}/{r}\right)\,,
\ee
to large $r$. Using the asymptotic behavior in (\ref{say}) one finds:
\be
R^{ext}_{\ell m}(r)\to { X\over W}\big(C\, r^{h-1}+D\, r^{-h}\big) \qquad \textrm{for $r\to \infty$}\, .
\ee
It follows that
\be
J_{\ell m}=\frac{ X}{W}\,C\,,
\ee
and the perturbed action may be written as
\be S = S_{CFT}+\sum_{\ell, m}\int dt^+dt^- J_{\ell m }e^{im (t^++3r_0t^-/4)} \mathcal{O}_{\ell }(t^+, t^-)\,.\ee

This perturbation will induce transitions out of or `decays' of the vacuum state.
Fermi's golden rule then gives the transition rate \cite{hep-th/9702015, hep-th/9706100}:
\be\label{total rate}
\mathcal{R}=2\pi \sum_{\ell, m}|J_{\ell m}|^2 \int dt^+ dt^-  e^{-imt^+- im3r_0t^-/4}G(t^+,t^-)\,.
\ee
Here  $G(t^+,t^-)=\langle \mathcal{O}^\dagger(t^+,t^-)\mathcal{O}(0,0)\rangle_{T_L}$ is the two point function of the dual two-dimensional conformal field theory, which has a finite left-moving temperature $T_L={1/2\pi}$ and angular potential. Such two point functions are fixed up to an overall normalization  $C_{\mathcal{O}}^2$ by conformal invariance. Precisely  these two point functions were studied in the context of superradiant scalar scattering in \cite{0907.3477}, to which we refer the reader for details. One finds, with the appropriate $i\epsilon$ prescription, the general formula:
\be
\int dt^+dt^-e^{-i\omega_Lt^+-i\omega_Rt^-}G(t^+,t^-)=C_{\mathcal{O}}^2\frac{(2\pi T_L)^{2h_L-1}}{\Gamma(2h_L)} e^{-\omega_L/2T_L} \left|\Gamma\left(h_L+i\frac{\omega_L}{2\pi T_L}\right)\right|^2 \frac{2\pi}{\Gamma(2h_R)}\omega_R^{2h_R-1}\,.
\ee
Substitution into (\ref{total rate}) then gives:
\be\label{partial wave rate}
\mathcal{R}_{\ell m}=C_\mathcal{O}^2\, \frac{(2\pi)^2(3r_0/4)^{2h-1}}{\Gamma(2h)^2}\, |J_{\ell m}|^2\, m^{2h-1} e^{-\pi m}|\Gamma(h+im)|^2\,  \,,
\ee
for $m>0$.

\subsection{Comparison}
The holographic dictionary predicts that the vacuum decay rate in the CFT equals the particle flux across the horizon.
Indeed we see that for the normalization
\be \label{dd} C_{\mathcal O}= {2^{h-1}(2h-1)\over  2\pi}M\,,\ee
we have exactly:
\be
\mathcal{R}_{\ell m}={\lambda^2\over 4} r_0 m^{-1}e^{-\pi m}S_\ell^2(\pi/2)\frac{\left|\Gamma(h+im)\right|^2}{\Gamma(2h)^2} \left|M_{im,h-\half}\left({3im}/{2}\right)\right|^2={\cal F}_{\ell m}\,.
\ee
Hence, up to the normalization (\ref{dd}),  we find agreement between the gravity and CFT computations.

Recent investigations of Kerr/CFT suggest that the dual CFT may have some exotic features such as nonlocality \cite{ElShowk:2011cm, Song:2011sr, Azeyanagi:2012zd} or a single Virasoro-Kac-Moody in place of two Virasoros \cite{Detournay:2012pc, Compere:2013aya}. Given this one may wonder why the very detailed agreement found here, which rests on the standard CFT formulae for correlators and deformations, works so well.  One possibility is simply that the standard CFT formulae are highly universal and remain valid for the exotic variations  under consideration. We leave this to future investigations.

\subsection{Reattaching the asymptotically flat region}

In this subsection we reattach the asymptotically flat region, and compute the flux of scalar radiation at future null infinity generated by a star on the orbit \eqref{plunge in NHEK} coupled to a scalar via the interaction \eqref{Sint}.
Expanding the scalar field in modes
\be
\Psi_{\ell m \hat\omega}=e^{-i\hat\omega \hat t}e^{im\hat\phi}\hat S_\ell(\theta) \hat R_{\ell m \hat\omega}(\hat r)\,,\label{psi expansion}
\ee
the wave equation in the full extreme Kerr separates into
\begin{eqnarray}
\frac{d}{d\hat r}\left(\Delta\frac{d \hat R_{\ell m \hat\omega}}{d\hat r} \right) + \left({H^2\over\Delta} + 2 M m \hat\omega -\hat K_\ell\right)\hat R_{\ell m \hat\omega}&=&\hat T_{\ell m \hat\omega}\,,\label{radial eqn Kerr} \\
{1\over\sin\theta}\p_\theta(\sin\theta\,\p_\theta \hat S_\ell)+\left(\hat K_\ell-\frac{m^2}{\sin^2\theta}-M^2\hat\omega^2\sin^2\theta\right)\hat S_\ell&=&0\,,\label{angular eqn Kerr}
\end{eqnarray}
where $H=(\hat r^2+M^2)\hat\omega-Mm\,,$ and $\hat T_{\ell m \hat\omega}$ is the source term for the orbiting star.

In the coordinates \eqref{coord transn} the radial equation \eqref{radial eqn Kerr} becomes
\be\label{radial x eqn}
\p_r(r^2\p_r\hat R_{\ell m \hat\omega}) + V \hat R_{\ell m \hat\omega} = \hat T_{\ell m \hat\omega}\,,
\ee
with
\be
V =  \left({M\hat\omega r^2 + 2 M \hat\omega r + k/2 \over r}\right)^2 + 2Mm\hat\omega - \hat K_\ell \,,\qquad k=4M(\hat\omega-{m\over 2M})\,.
\ee

The orbiting star sources the scalar deep in the NHEK region at small $r$. Generic modes do not penetrate into this region because the potential $V$ diverges like $ 1 / r$. However when the frequency is near the superradiant bound, i.e. when
\be
k\equiv2\omega=-{3\over 2}m r_0\,, \quad\mathrm{with}\quad r_0\ll 1\,,
\ee
the divergence is absent and the modes penetrate.
We then have to leading order:
\be\label{relations for k<<1}
2M\hat\omega=m\,, \quad \hat K_\ell=K_\ell \,, \quad \hat S_\ell=S_\ell \,, \quad 2M\hat\omega-m=k/2\,.
\ee
In the near region $r\ll 1$ the modes \eqref{psi expansion} agree with the ones in \eqref{psi simpler  expansion} and equation \eqref{radial x eqn} becomes the NHEK equation \eqref{radial eqn} whose solutions were studied in section 3.1. In the far region $r\gg k$ equation \eqref{radial x eqn} becomes
\be
\p_r(r^2\p_r\hat R_{\ell m \hat\omega}) + \left[{m^2\over 4}(r+2)^2+m^2-K_\ell\right] \hat R_{\ell m \hat\omega}=0\,.
\ee
The solution of this with no incoming flux at past null infinity is
\be
\hat R_{\ell m \hat\omega}^{far}(r)=P\, r^{h-1}e^{-imr/2}{_1}{F}{_1}\left(h+im\,,2h\,,imr\right)+ Q\,(h\to 1-h)\,,
\ee
where
\be\label{P/Q}
{P\over Q}=-(-im)^{2h-1}\frac{\Gamma(2-2h)}{\Gamma(2h)}\frac{\Gamma(h-im)}{\Gamma(1-h-im)}\,.
\ee
The asymptotic behaviors are:
\begin{eqnarray}
\hat R_{\ell m \hat\omega}^{far}(r\to \infty)&=&Q\,\frac{\Gamma(2-2h)}{\Gamma(1-h+im)}(im)^{h-1+im}\times\\
 &&~~~\times\left[1- \frac{(-im)^{2h-1}}{(im)^{2h-1}}\frac{\sin\pi(h+im)}{\sin\pi(h-im)}\right]r^{-1+im}e^{imr/2} \,,\notag\\
\hat R_{\ell m \hat\omega}^{far}(r\to 0)&=&P\, r^{h-1}+Q\,r^{-h}\,.\label{R^far behaviors}
\end{eqnarray}

The demand of no incoming flux from null infinity does not fix the overall magnitude of the solution.
This is fixed by matching it with a solution of \eqref{radial eqn}, which includes the near-region source, with no incoming flux at the past horizon. We use the method of matched asymptotic expansions as in \cite{0907.3477,Teukolsky:1974yv}, matching the small $r$ behavior of $\hat R_{\ell m \hat\omega}^{far}$ above with the large $r$ behavior of a solution of \eqref{radial eqn} from section 3.1. This specifies the magnitude of the solution in terms of the source according to:
\be
Q=\frac{ Z}{W} (-2i\omega)^{h}\left[1-(3r_0/2)^{2h-1}|m|^{4h-2} \frac{\Gamma(1-2h)^2}{\Gamma(2h-1)^2}\frac{\Gamma(h-im)^2}{\Gamma(1-h-im)^2}\right]^{-1}\,.
\ee
Putting this all together we have finally the outgoing radiation flux at future null infinity:
\bea
\frac{d\hat E}{d\hat t}&=&{\lambda^2\over 24M^2}S_\ell^2(\pi/2)\left|W_{im,h-\half}(3im/2)\right|^2 (3r_0/2)^{2\mathrm{Re}[h]} m^{4\mathrm{Re}[h]-2}e^{\pi m}\times \\
&&\qquad\times  \frac{|2h-1|^2  |\Gamma(h-im)|^4 /|\Gamma(2h)|^4}{\left|1-(3r_0/2)^{2h-1}m^{4h-2} \frac{\Gamma(1-2h)^2}{\Gamma(2h-1)^2}\frac{\Gamma(h-im)^2}{\Gamma(1-h-im)^2}\right|^2}\,, \notag
\eea
for $m>0$. 

Although we shall not attempt to do so in this paper, the above result should also be obtainable from a yet-more-refined CFT analysis. Imposing Dirichlet or Neumann conditions for real $h$ at the NHEK boundary ensures there is no energy flux through the boundary and decouples the CFT from the far region. When the far region is attached, we of course want to allow flux to leak out of the NHEK region, but in a very specific way that gives only outgoing flux at null infinity. This requires ``leaky'' boundary conditions in which the fast and slow modes - or sources and vevs in the CFT - are related in amplitude by the ratio $P/Q$ in equation \eqref{P/Q}. $P$ and $Q$ are determined by the physics of the far region and cannot be determined either by a bulk NHEK or boundary CFT calculation. An alternate approach to a CFT analysis would be to compute amplitudes directly  rather than the probabilities given by their squares. The problem then becomes linear in the modes and the bulk result will just be a linear sum of CFT amplitudes with Dirichlet/Neumann boundary conditions weighted by $P$ and $Q$.

\section{Gravitational radiation from a star orbiting near the horizon}
In this section we  generalize the previous scalar analysis to the gravity case.
\subsection{Gravity computation in NHEK}
Gravity waves are most efficiently analyzed in the Newman-Penrose formalism \cite{Chandra} which involves the null tetrad $(l,n,m,\bar m)$.  Teukolsky showed \cite{PhysRevLett.29.1114, Teukolsky:1973ha} that the perturbation of the Weyl scalar
\be \psi_4= C_{\alpha\beta\gamma\delta}n^\alpha\bar m^\beta n^\gamma\bar m^\delta\,, \ee
in any Type D background obeys:
\be\label{Teukolsky psi4 eqn}
[(\Delta+3\gamma-\bar\gamma+4\mu+\bar\mu)(D+4\epsilon-\rho)-(\bar\delta-\bar\tau+\bar\beta+3\alpha+4\pi)(\delta-\tau+4\beta)-3\psi_2]\delta\psi_4=4\pi T_4\,,
\ee
where,
\begin{eqnarray}\label{T4}
T_4&=&(\Delta+3\gamma-\bar\gamma+4\mu+\bar\mu)[(\bar\delta -2\bar\tau+2\alpha)T_{n\bar m}-(\Delta+2\gamma-2\bar\gamma-\bar\mu) T_{\bar m \bar m}]\nonumber\\
&&+(\bar\delta-\bar\tau+\bar\beta+3\alpha+4\pi)[(\Delta+2\gamma+2\bar\mu)T_{n\bar m}-(\bar\delta-\bar\tau+2\bar\beta+2\alpha)T_{nn}]\,.
\end{eqnarray}
Here $D=l^\mu\p_\mu\,,\Delta=n^\mu\p_\mu\,,\delta=m^\mu\p_\mu\,,\bar\delta=\bar m^\mu\p_\mu$ and $T_{n\bar m}=T_{\mu\nu}n^\mu\bar m^\nu\,,$ etc. For NHEK we choose the Kinnersley tetrad 
($l\cdot n=-m\cdot \bar{m}=-1$):
\begin{eqnarray}
l^\mu &=&\left(\frac{1}{r^2},1,0,-{1\over r}\right)\,,\notag\\
n^\mu &=&\frac{1}{2M^2(1+\cos^2\theta)}\left(1,-r^2,0,-r\right)\,,\notag \\
m^\mu &=&\frac{1}{\sqrt{2}M(1+i\cos\theta)}\left(0,0,1,\frac{i(1+\cos^2\theta)}{2\sin\theta}\right)\,,\label{NHEK NP tetrad}
\end{eqnarray}
so that the nonzero spin coefficients are:
\begin{eqnarray}\label{NP coeffs}
&&\tau =-\frac{i\eta\bar{\eta}\sin\theta}{\sqrt{2}M}\,, \quad \pi =\frac{i\eta^2\sin\theta}{\sqrt{2}M}\,, \quad
\beta =-\frac{\bar{\eta}\cot\theta}{2\sqrt{2}M}\,,\nonumber \\  &&\qquad\qquad\qquad \alpha=\pi-\bar{\beta}\,, \quad \gamma =\frac{\eta\bar{\eta}r}{2M^2}\,,
\end{eqnarray}
where,
\be
\eta\equiv-\frac{1}{1-i\cos\theta}\,,
\ee
and the Weyl scalars are: $\psi_0=\psi_1=\psi_3=\psi_4=0\,,\psi_2=\eta^3/M^2$. The energy-momentum tensor for a point particle of rest mass $m_0$ on a geodesic $x^\alpha_*(\tau)$ is given by
\be
T^{\mu\nu}=m_0\int d\tau (-g)^{-1/2} \frac{dx^\mu}{d\tau}\frac{dx^\nu}{d\tau}\delta^{(4)}(x^\alpha-x^\alpha_*(\tau))\,,
\ee
which for our simple geodesic \eqref{plunge in NHEK} in NHEK has only one non-vanishing component (setting $\phi_0=0$):
\be\label{stress tensor}
T_{\phi\phi}=\frac{m_0 r_0}{\sqrt{3}M}\delta(r-r_0)\delta(\theta-{\pi\over 2})\delta(\phi+{3\over 4}r_0 t)\,.
\ee
Then from \eqref{T4} we find:
\begin{eqnarray}\label{T4 for NHEK plunge}
T_4&=&\frac{m_0r_0^3}{64\sqrt{3}M^7}\Big[ 144 \, \delta(r-r_0)\delta(\theta-{\pi\over 2})\delta(\phi+{3\over 4}r_0 t) +\,16\, r_0\delta'(r-r_0)\delta(\theta-{\pi\over 2})\delta(\phi+{3\over 4}r_0 t)\nonumber\\
&&\qquad -\,48i\, \delta(r-r_0)\delta'(\theta-{\pi\over 2})\delta(\phi+{3\over 4}r_0 t) -\,21\, \delta(r-r_0)\delta(\theta-{\pi\over 2})\delta'(\phi+{3\over 4}r_0 t)\nonumber\\
&&\qquad -\,8i\, r_0\delta'(r-r_0)\delta'(\theta-{\pi\over 2})\delta(\phi+{3\over 4}r_0 t) -\,3\, r_0\delta'(r-r_0)\delta(\theta-{\pi\over 2})\delta'(\phi+{3\over 4}r_0 t)\nonumber\\
&&\qquad +\,6i\, \delta(r-r_0)\delta'(\theta-{\pi\over 2})\delta'(\phi+{3\over 4}r_0 t) +\,2\, r_0^2\delta''(r-r_0)\delta(\theta-{\pi\over 2})\delta(\phi+{3\over 4}r_0 t)\nonumber\\
&&\qquad -\,8\, \delta(r-r_0)\delta''(\theta-{\pi\over 2})\delta(\phi+{3\over 4}r_0 t) +\,{9\over 8}\, \delta(r-r_0)\delta(\theta-{\pi\over 2})\delta''(\phi+{3\over 4}r_0 t)\Big]\,.
\end{eqnarray}
Equation \eqref{Teukolsky psi4 eqn} separates in NHEK for the variable:
\be\label{psi(-2) def}
\psi^{(-2)}\equiv\eta^{-4}\delta\psi_4\,.
\ee

As in the scalar case, we respect the symmetry \eqref{KIlling vector} and construct a solution using the expansion
\begin{eqnarray}
\psi^{(-2)}&=&\sum_{\ell,m}e^{im\left(\phi+3r_0t/4\right)}S_\ell(\theta) R_{\ell m}(r)\,,\label{psi simpler  expansion s}\\
4\pi{\cal T}&=&\eta\bar\eta \sum_{\ell,m}e^{im\left(\phi+3r_0t/4\right)}S_\ell(\theta) T_{\ell m }(r)\,,\label{source simpler expansion s}
\end{eqnarray}
where ${\cal T}\equiv 2M^2\eta^{-4} T_4$ and $S_\ell$ are now the spin-weighted spheroidal harmonics obeying
\be
{1\over\sin\theta}\p_\theta(\sin\theta\,\p_\theta S_\ell)+\left(K_\ell-\frac{m^2+s^2+2ms\cos\theta}{\sin^2\theta}- \frac{m^2}{4}\sin^2\theta-ms\cos\theta\right)S_\ell=0\,,\label{angular eqn s=-2}
\ee
with $K_\ell$ a separation constant, $\ell\ge|s|\,, -\ell\leq m\leq \ell$. In general, for gravity waves we have $s=\pm 2$ but in particular, since we have chosen to work with the Weyl scalar $\psi_4$, for us:
\be s=-2\,. \ee
However, for convenience and because some (though not all) of the formulae that follow apply to more general spin we often leave $s$ in the equations. As before, $S_\ell$ and $K_\ell$ depend on $m$ and $s$ as well but we write only their primary index $\ell$ to avoid clutter. We also assume the same normalization \eqref{sph harmonics normn}. The expansion coefficients for $\mathcal{T}$ are
\begin{eqnarray}
T_{\ell m}(r)
&=&a_0\delta(r-r_0)+a_1 r_0\delta'(r-r_0)+a_2 r_0^2\delta''(r-r_0)\,,
\end{eqnarray}
where
\begin{eqnarray}
a_0&=&\frac{m_0r_0^3}{16\sqrt{3}M^5}\left(40S+3imS-{9\over 8}m^2S+6mS'-16iS'-8S''\right)\,,\notag\\
a_1&=&\frac{m_0r_0^3}{16\sqrt{3}M^5}\left(8iS'-3imS-16S\right)\,,\notag\\
a_2&=&\frac{m_0r_0^3}{16\sqrt{3}M^5}2S\,,
\end{eqnarray}
and $S, S', S''$ are $S_\ell(\pi/2), S'_\ell(\pi/2), S''_\ell(\pi/2)$ respectively. The separated radial equation becomes:
\be
r^{-2s}\p_r(r^{2s+2}\p_r R_{\ell m})+V_s R_{\ell m}=T_{\ell m}\,,\label{radial eqn s}
\ee
where $\omega$ is given in \eqref{Omega defn} and
\be
V_s(r)= 2m^2-K_\ell+s(s+1)+\frac{2\omega(m-is)}{r}+\frac{\omega^2}{r^2}\,.
\ee

This radial equation is again a Sturm-Liouville problem that is solved, for given boundary conditions, via standard Green's function methods. Two linearly independent solutions to the homogeneous radial equation ($T_{\ell m}=0$) are given by \cite{0906.2376}:
\be\label{Whittaker solns s}
\mathcal{M}(r)\equiv r^{-s}\,M_{im+s, h-\half}\left(-{2i\omega}/{r}\right)\,,\quad \mathcal{W}(r)\equiv r^{-s}\,W_{im+s, h-\half}\left(-{2i\omega}/{r}\right)\,,
\ee
where $h$ is given by \eqref{h defn}. These have simple behaviors at infinity and the horizon respectively:
\begin{eqnarray}
\mathcal{M}(r) &\to& (-2i\omega)^{h} r^{-h-s} \qquad\qquad\quad\, \textrm{for $r\to \infty$}\,,\\
\mathcal{W}(r) &\to& (-2i\omega)^{im+s} r^{-im-2s} e^{i\omega/r} \quad \textrm{for $r\to 0$}\,.
\end{eqnarray}
The corresponding solution of the radial equation \eqref{radial eqn s} is given by:
\be\label{radial soln s=-2}
R_{\ell m}(r)={1\over r_0^{-2s} W} \left[\mathcal{X}\,\Theta(r_0-r) \mathcal{W}(r)+ \mathcal{Z}\, \Theta(r-r_0) \mathcal{M}(r)\right]+a_2\delta(r-r_0)\,,
\ee
where $W$ is the $r$-independent Wronskian of the two solutions \eqref{Whittaker solns s},
\be
W= 2i\omega\frac{\Gamma(2h)}{\Gamma(h-im-s)}\,,
\ee
and
\bea
\mathcal{X}&=& r_0\mathcal{M}'(r_0)(2sa_2-a_1-2a_2)+\mathcal{M}(r_0)(a_0-2sa_1-2sa_2+4s^2a_2-a_2V_s(r_0))\,,\quad\cr
\mathcal{Z}&=&\mathcal{X}(\mathcal{M}\to\mathcal{W})\,.
\eea
Putting everything together we have:
\begin{eqnarray}
\psi^{(-2)}(r\to 0)&=&\sum_{\ell,m} e^{im(\phi+3r_0t/4)}\,S_\ell(\theta)\,\frac{\mathcal{X}}{r_0^{4}W} (-2i\omega)^{im-2}\, r^{-im+4}e^{-3imr_0/4r}\,,\\
\psi^{(-2)}(r\to \infty)&=&\sum_{\ell,m}e^{im(\phi+3r_0t/4)}\,S_\ell(\theta)\,\frac{\mathcal{Z}}{r_0^{4}W} (-2i\omega)^{h} \,r^{-h+2}\,.
\end{eqnarray}
The graviton number flux at the horizon is given by:
\bea
\mathcal{F}_{\ell m}&=&\frac{12 M^{10} r_0}{|\mathcal{C}|^2}m e^{-\pi m}\left|\frac{\mathcal{X}}{r_0^{4}W}\right|^2\,,\\
|{\mathcal C}|^2&\equiv &\left((K_\ell-m^2)^2+m^2\right)\left((K_\ell-m^2-2)^2+9m^2\right)\,,\nonumber
\eea
for $m>0$.

\subsection{Matching with CFT analysis}

The matching of the bulk gravity wave analysis with the boundary CFT analysis proceeds as in the scalar case with two minor differences. Again, in this subsection we assume $h$ is real.

The first difference arises because the perturbation $\delta\psi_4$ studied involves the second derivative of the metric, while the sources of the dual CFT are related to the falloff coefficients of the metric perturbation itself.
The latter (in ingoing radiation gauge) is obtained from a Hertz potential $\Psi_H$ as follows \cite{Chrzanowski:1975wv, Wald:1978vm, 0906.2380, 0908.3909}:
\be
h_{\mu\nu}=\Theta_{\mu\nu}\Psi_H\,,
\ee
where
\bea
\Theta_{\mu\nu}&=&-l_\mu l_\nu(\bar\delta+\alpha+3\bar\beta-\bar\tau)(\bar\delta+4\bar\beta+3\bar\tau) - \bar m_\mu \bar m_\nu (D-\bar\rho)(D+3\bar\rho)\notag\\
&&+l_{(\mu}\bar m_{\nu)}\left[(D+\rho-\bar\rho)(\bar\delta+4\bar\beta+3\bar\tau)+ (\bar\delta+3\bar\beta-\alpha-\pi-\bar\tau)(D+3\bar\rho)\right]\,.
\eea
The Hertz potential itself solves the $s=-2$ vacuum Teukolsky equation and is given by:
\be
\Psi_H={4\over\mathcal{C}}R^{(-2)}(r)S^{(+2)}(\theta)e^{-i\omega t+im\phi}\,,
\ee
where
\bea
&&R^{(-2)}(r)={\mathcal{X}\over r_0^{-2s} W} \mathcal{W}(r)\\
&&~~~~\to {\mathcal{X}\over r_0^{-2s} W} \left[\frac{(-2i\omega)^{1-h}\Gamma(2h-1)}{\Gamma(h-im-s)}r^{h-1-s}+ \frac{(-2i\omega)^h\Gamma(1-2h)}{\Gamma(1-h-im-s)}r^{-h-s}\right]\quad \textrm{for $r\to \infty$}\,.\notag
\eea
Reading off the CFT source from the leading term of the Hertz potential at the boundary we get:
\be
J_{\ell m}={4\over \mathcal{C}}{\mathcal{X}\over r_0^{-2s} W} \frac{(-2i\omega)^{1-h}\Gamma(2h-1)}{\Gamma(h-im-s)}\,.
\ee

The second difference is that instead of $h_L=h_R=h$, the Kerr CFT dictionary \cite{0908.3909} implies:
\be
h_R=h\,,\quad h_L=h-s\,.
\ee
Normalizing the operators with
\be
{C_{\cal O}}=\frac{2^{h-1}M^5}{2\pi}\frac{\sqrt{\Gamma(2h+4)\Gamma(2h)}}{\Gamma(2h-1)}\,,
\ee
we again find agreement between the bulk and boundary computations.

\subsection{Reattaching the asymptotically flat region}

In this subsection we compute the flux of gravitational radiation out to future null infinity sourced by a star in a near horizon orbit in extreme Kerr. As in the scalar case, the radiation is dominated by near-superradiant modes, because the effective potential prohibits other modes from penetrating to the source.
Using the near-superradiant relations \eqref{relations for k<<1}, in the near region $r\ll 1$ the full spin-two Teukolsky equation for $\hat\psi^{(-2)}\equiv(\hat r-iM\cos\theta)^4\delta\hat\psi_4$ reduces to the NHEK equation  \eqref{radial eqn s} and in the far region $r\gg k$ it becomes
\be
r^{-2s}\p_r(r^{2s+2}\p_r\hat R_{\ell m \hat\omega}) + \left[{m^2\over 4}(r+2)^2+ismr+m^2+s(s+1)-K_\ell\right] \hat R_{\ell m \hat\omega}=0\,.
\ee
The solution of this that is purely outgoing at null infinity (i.e. behaves as $r^{-1+im-2s}e^{imr/2}$ for large $r$) is given by:
\be
\hat R_{\ell m \hat\omega}^{far}(r)=P\, r^{h-1-s}e^{-imr/2}{_1}{F}{_1}\left(h+im-s\,,2h\,,imr\right)+ Q\,(h\to 1-h)\,,
\ee
with,
\be\label{P/Q s}
{P\over Q}= -(-im)^{2h-1}\frac{\Gamma(2-2h)}{\Gamma(2h)}\frac{\Gamma(h-im+s)}{\Gamma(1-h-im+s)}\,.
\ee
Asymptotically:
\begin{eqnarray}
\hat R_{\ell m \hat\omega}^{far}(r\to \infty)&=&Q\,\frac{\Gamma(2-2h)}{\Gamma(1-h+im-s)}(im)^{h-1+im-s}\times\\
 &&~~~\times\left[1- \frac{(-im)^{2h-1}}{(im)^{2h-1}}\frac{\sin\pi(h+im)}{\sin\pi(h-im)}\right]r^{-1+im-2s}e^{imr/2} \,,\notag\\
\hat R_{\ell m \hat\omega}^{far}(r\to 0)&=&P\, r^{h-1-s}+Q\,r^{-h-s}\,.\label{R^far behaviors s}
\end{eqnarray}
$Q$ is again fixed by matching with a solution of \eqref{radial eqn s} which is purely ingoing at the horizon. One thereby finds (note that $\hat\psi^{(-2)}=M^6 \psi^{(-2)}$):
\bea
Q&=&M^6 \frac{\mathcal{Z}}{r_0^{-2s}W} (-2i\omega)^{h}\times\\
 &&~~~\times\left[1-(3r_0/2)^{2h-1}|m|^{4h-2} \frac{\Gamma(1-2h)^2}{\Gamma(2h-1)^2} \frac{\Gamma(h-im-s)\Gamma(h-im+s)}{\Gamma(1-h-im-s)\Gamma(1-h-im+s)}\right]^{-1}\,.\notag
\eea
Finally this implies the outgoing energy flux at future null infinity, for $m>0$, is:
\bea
\frac{d\hat E}{d\hat t}&=& {m_0^2\over 2^5 3^3 M^2}|\gamma_{\ell m}|^2 (3r_0/2)^{2\mathrm{Re}[h]} m^{4\mathrm{Re}[h]-2}e^{\pi m} \times  \notag\\
&&\quad\times \frac{|2h-1|^2  |\Gamma(h+2-im)|^2 |\Gamma(h-2-im)|^2/|\Gamma(2h)|^4}{\left|1-(3r_0/2)^{2h-1}m^{4h-2} \frac{\Gamma(1-2h)^2}{\Gamma(2h-1)^2} \frac{\Gamma(h-im-s)\Gamma(h-im+s)}{\Gamma(1-h-im-s)\Gamma(1-h-im+s)}\right|^2}\,,
\eea
where
\bea
\gamma_{\ell m}&=&2\left[(h^2-h+6-im)S+4(2i+m)S'-4S''\right]W_{im-2,h-\half}(3im/2)\notag\\
&&+\left[(4+3im)S-8iS'\right]W_{im-1,h-\half}(3im/2)\,.
\eea
For example, it is expected \cite{Detweiler:1978ge} that  an important contribution comes from the $\ell=m=2$ mode. For this mode the eigenvalue $K_\ell$ has been computed numerically to good accuracy in \cite{Press:1973zz} and corresponds to $h=0.5+2.05i$. The eigenfunction $S_\ell(\theta)$ can then be computed numerically to good accuracy as well using Leaver's series solution as described in \cite{Leaver:1985ax, gr-qc/0511111}. We find:
\be
\frac{d\hat E}{d\hat t}=0.10\,{m_0^2\over M^2}r_0\,.
\ee

%

\section*{Acknowledgements}

We are grateful to M. Guica, S. Hadar, T. Hartman, J. Hewlett and W. Song for useful conversations. This work was supported in part by DOE grant DE-FG02-91ER40654.

\end{document}